# COLOR TRANSPARENCY

N.N.NIKOLAEV[1,2] and B.G.ZAKHAROV[2]

[1] *Institut f. Kernphysik, Forschungszentrum Jülich GmbH*
*D-52425 Jülich, Germany*

and

[2] *L.D.Landau Institute for Theoretical Physics, GSP-1, 117940, ul.Kosygina 2*
*V-334 Moscow, Russia*

## ABSTRACT

The current status of the theory of and the experimental evidence for color transparency are reviewed. The problems with interpretation of quasielastic scattering on nuclei are discussed to some detail.

## 1. What is color transparency?

Intranuclear absorption of the incoming and outgoing particles makes nuclear production cross sections $\sigma_A$ smaller than $A$ times the free nucleon cross section $\sigma_N$. For instance, incoherent photoproduction of the $\rho^0$ on lead is suppressed by one order in the magnitude. For color transparency (CT) phenomenon [1, 2], much weaker suppression is expected in virtual photoproduction (leptoproduction) $\gamma^* A \to \rho^0 A^*$ in deep inelastic scattering (DIS) at large $Q^2$ and high energy $\nu$. How is this possible?

Leptoproduction of vector mesons illustrates nicely CT ideas: i) virtual photons shrink with $Q^2$ [3], ii) absorption of the photon releases a color singlet ejectile system of quarks (and gluons) of small transverse size and small interaction cross section, iii) for the Lorentz dilation of time the ejectile retains its small size during the propagation in a nucleus entailing weak intranuclear absorption. The key QCD ingredients are the small-size dominated mechanism of the hard reaction and the weak absorption of small size color singlet states. CT tests of QCD are widely discussed as one of the motivations for the new 15-30 GeV electron facility - ELFE (European Laboratory For Electrons) [4, 5], some tests of CT can be performed at CEBAF [6].

The key to CT physics is a $q\bar{q}$ color dipole description of high energy hadrons and photons with the dipole size $r$ being the $q\bar{q}$ separation in the transverse plane [1, 3] ($3q$ baryons are a system of 3 color dipoles). The flavour blind pomeron dipole cross section $\sigma(\nu, r)$ unifies description of the different processes. (For the color dipole picture of inclusive DIS see [7]). The conditions for CT are met in different reactions to a varying accuracy and degree of certainty. In this talk we review the current status of the theory of CT and the experimental evidence for CT. In many cases we are bound to fleeting remarks, for extensive reviews see [4]-[6],[8]-[10].

## 2. What do we know about the color dipole cross section?

Exclusive vector meson production $\gamma^* p \to V p$ is the typical pomeron exchange diffractive reaction. In [11, 9, 12] it was shown that because of a shrinkage of the $q\bar{q}$ Fock state of the photon at large $Q^2$ and/or large mass of heavy quarks [3], the vector meson production amplitudes $\mathcal{M}_{T,L} = \langle V_{T,L} | \sigma(\nu, r) | \gamma^*_{T,L} \rangle$ are dominated by the contribution from dipoles of size $r \sim r_S$, where the scanning radius $r_S = 6/\sqrt{m_V^2 + Q^2}$. For the transverse (T) and longitudinally (L) polarized mesons one finds amplitudes of the form [12, 13] $\mathcal{M}_T \propto r_S^2 \sigma(\nu, r_S)$, $\mathcal{M}_L \approx \frac{\sqrt{Q^2}}{m_V} \mathcal{M}_T$. Inverting the problem, one can deduce $\sigma(\nu, r)$ from the experimental data on real and virtual photoproduction. Such an analysis was done recently with the the results [14] shown in Fig. 1. There is a clear evidence for a roughly $\propto r^2$ CT decrease of $\sigma(\nu, r)$ towards small dipole size $r$. Notice a consistency of $\sigma(\nu, r)$ determined from the $\rho^0, \phi^0$ and $J/\Psi$ production (there remain some uncertainties with the poorly known radius and wave function of $\rho^0, \phi^0$). The same color dipole cross section gives an excellent unified description of the structure function of DIS on protons and of diffractive DIS at HERA [15].

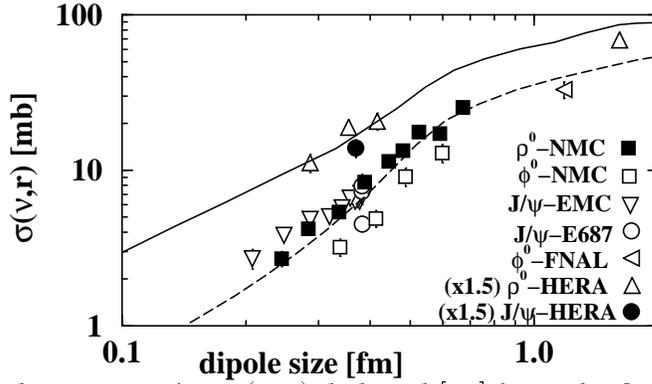

Figure 1: The dipole cross section $\sigma(\nu, r)$ deduced [14] from the fixed-target ($\nu \sim 100$-$300$ GeV) and HERA ($\nu \sim 10^4$ GeV) data on vector meson production. The solid and dashed curves are the corresponding predictions from the color dipole BFKL dynamics [15, 13].

### 3. CT in electroproduction of vector mesons off nuclei.

Important scales are the nuclear radius $R_A$, the interaction length $l_{int} = 1/n_A \sigma_{tot}(VN)$, the distance from the target at which the photon fluctuates into the partonic state (the coherence length) $l_c = 2\nu/(Q^2 + m_V^2)$ and the scale of recombination of ejectile partons into the observed vector meson (the formation length) $l_f = \frac{\nu}{m_V \Delta m}$, where $\Delta m$ is the typical level splitting in the quarkonium. Typically, $l_c \ll l_f$. Absorption of the photon produces the $q\bar{q}$ ejectile state. Any $q\bar{q}$ wave packet can be expanded in the complete set of hadronic eigenstates and, when done correctly, the partonic and hadronic basis descriptions are equivalent. The partonic basis is convenient because of the exact high-energy diagonalization of the scattering matrix in the color dipole representation, the hadronic basis is convenient for the description of temporal coherency of the ejectile wave packet at subasymptotic energies. The formation length $l_f$ is a scale of the phase decorrelation of different hadronic components of the ejectile. At low energies $l_f \ll l_{int}, R_A$, decorrelation is fast and strong attenuation with the free nucleon cross section is found. At high energies such that $l_f \gtrsim l_{int}, R_A$, electroexcited higher mass states (radial excitations $nS,...$) regenerate the observed vector meson $V$ by intranuclear interaction and it is the coherency of the direct and regenerated $V$s which is the underlying mechanism of CT.

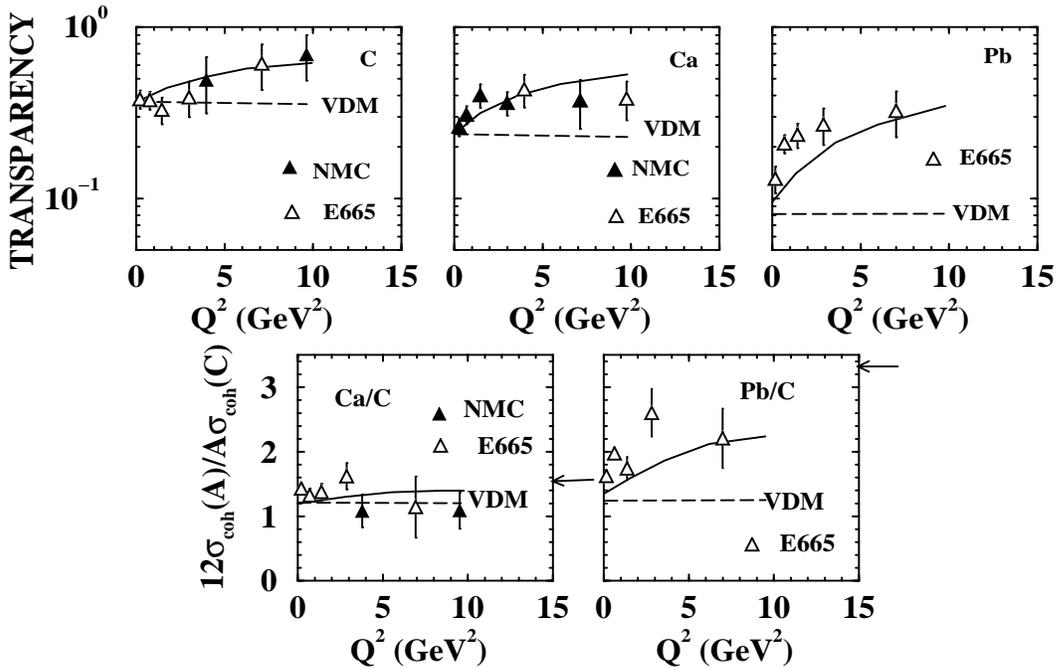

Figure 2: The $Q^2$ dependence (top) of nuclear transparency $T_A = \sigma_A/A\sigma_N$ in incoherent production of the $\rho^0$ and (bottom) of the ratio of coherent production cross sections per bound nucleon [16, 17]. The solid curves are predictions of the color dipole model of CT [12], shown also are the VDM predictions [18] and (arrows) the coherent cross section ratio for the vanishing intranuclear attenuation.

The E665 [16] and NMC [17] results on the $\rho^0$ production are shown in Fig. 2 and confirm the predicted [12] decrease of nuclear attenuation with $Q^2$, although the error bars are still large. At $Q^2 \sim 10\,\text{GeV}^2$ the predicted and observed residual attenuation is only about twice weaker than at $Q^2 = 0$, which must be contrasted to the CT suppression of the free nucleon production amplitude by the factor $5-6$ over the same range of $Q^2$. The profound reason for this slow onset of CT is the very mechanism of CT, by which intranuclear rescatterings are selective to the dipole size which is substantially larger than $r_S$ [6, 9, 12].

Furthermore, the residual nuclear attenuation is still strong, whereas in inclusive DIS at the same Bjorken variable $x = (Q^2 + m_\rho^2)/2m_p\nu \sim 0.03$ nuclear shadowing is negligibly weak [19]. This subtle breaking of factorization theorems by nuclear shadowing has two origins [6, 20]. First, nuclear shadowing in exclusive vector meson production only requires $l_f \gtrsim R_A$ compared to an exceedingly stronger condition $l_c \approx \frac{m_V \Delta m}{Q^2 + m_V^2} l_f \gtrsim R$ in inclusive DIS on nuclei. Second, the dominant attenuation of $q\bar{q}$ Fock states of the photon is the leading twist in DIS and a higher twist in the vector meson production, although in the latter case the decrease with $Q^2$ is extremely slow. Eventually, at very large $Q^2$ the leading twist attenuation of the $q\bar{q}g$ Fock states of the photon and vector meson will take over. It can be cast in the form of nuclear modification of gluon densities which is different for exclusive and inclusive DIS on nuclei, though; this factorization breaking was not noticed in [21].

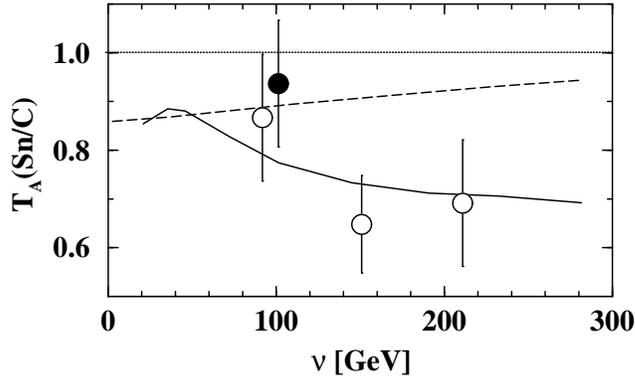

Figure 3: Nuclear transparency in incoherent $J/\Psi$ photoproduction vs. energy (NMC [23]). The solid curve is the prediction of the multiple-scattering theory of Ref. [22], the dashed curve is the prediction of the quasiclassical diffusion model of Ref. [24].

The rôle of the coherence length $l_c$ is somewhat special [11, 22, 9, 12]: if $l_c \gg R_A$ then production of vector mesons on different nucleons at the same impact parameter is coherent and the whole thickness of the nucleus contributes to attenuation of the $q\bar{q}$ pair compared to approximately half of nuclear thickness at $l_c \ll R_A$, when production on different nucleons is incoherent. The NMC data [23] shown in Fig. 3 confirm this prediction and rule out the "quantum" and "classical" diffusion model of [24] which predicts the decrease of nuclear attenuation with energy. This failure of the model [24] derives from its intrinsic inconsistency with the multiple scattering theory.

## 4. CT and electroproduction of radially excited vector mesons

The wave function of $2S$ radial excitation has a node. At $Q^2 \sim 0$ the scanning radius $r_S$ is close to the position $r_n \sim R_V$ of the node and there are strong cancellations (the node effect) of large size, $r \gtrsim r_n$, and small size, $r \lesssim r_n$, contributions to the production amplitude. This correctly explains [11, 9] the observed $\sigma(\Psi')/\sigma(J/\Psi) = 0.20 \pm 0.05(stat) \pm 0.07(syst)$ [25]. The larger is $Q^2$, the smaller is $r_S$ and the weaker are cancellations and $\sigma(2S)/\sigma(1S) \sim 1$ is predicted at $Q^2 \to \infty$ [26, 13, 14]. The first evidence for such a rise of the $\rho'/\rho^0$ production ratio from the E665 experiment [27] is shown in Fig. 4. (It is not yet clear which of the two $\rho'$ mesons is the $2S$ state.)

For production on nuclei, the node effect leads to a counterintuitive prediction of $T_A(2S) > T_A(1S)$, even to an antishadowing $T_A(2S) > 1$ [11, 9, 12, 26, 28], despite the obviously larger $V(2S)N$ cross section. The onset of antishadowing has a lively $(Q^2, \nu)$ dependence which can be observed at CEBAF (Fig. 5). Here our emphasis is on the prediction of the rapid variation of $T_A(2S)$ on the scale $Q^2 \lesssim 1\,\text{GeV}^2$ which derives from the rapid $Q^2$ dependence of the scanning radius. This entirely quantum-mechanical antishadowing effect is completely missed in the semiclassical expansion model [24].

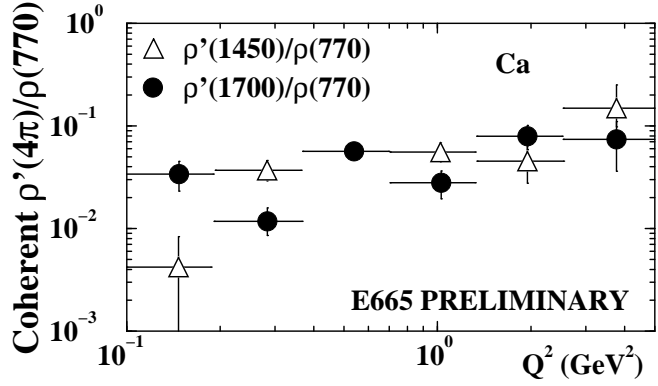

Figure 4: The E665 data on the $Q^2$ dependence of the $\rho'(4\pi)/\rho^0$ ratio in coherent leptoproduction on the $Ca$ nucleus [27].

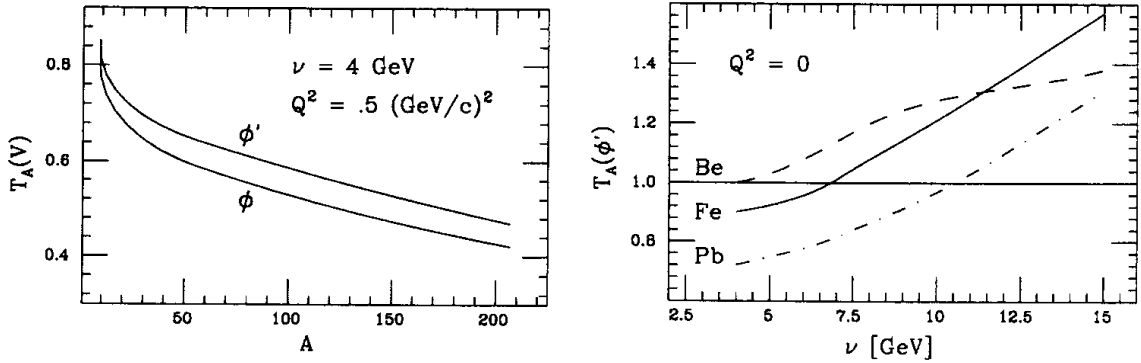

Figure 5: Target and energy dependence of nuclear transparency in leptoproduction of the $\phi'$ meson expected in the color dipole model of CT [28].

## 5. CT in the charge exchange of pions on nuclei.

The historically first evidence for CT effect in hadronic reactions was obtained by Kopeliovich and Zakharov [29] from the experimental data [30] on the pion CEX. The underlying quark mechanism of CEX $\pi^- p \to \pi^0 n$ is the CEX $\bar{u} \to \bar{d}$ by the $\rho$-reggeon exchange. The absolute magnitude of the reggeon exchange amplitude is insensitive, whereas the $t$-dependence of the amplitude is sensitive, to the $q\bar{q}$ separation in the pion. (Notice the contrast between the weak and strong dependence on the $q\bar{q}$ separation of the reggeon and pomeron exchanges, respectively). In CEX on a nucleus $\pi^- A \to \pi^0 A^*$, the large-size components of the incoming and outgoing pion waves are stripped off by intranuclear attenuation. The resulting enhancement of small $q\bar{q}$ separations leads to an unambiguous CT signal [29] - the rise of nuclear transparency with $t$, which brings the theory and the experiment together [30], see Fig. 6. Notice the very small error bars. Further experiments of this kind are worth while, because the cross sections are large and the interpretation of the results in terms of the size of $q\bar{q}$ states involved in the reaction is straightforward.

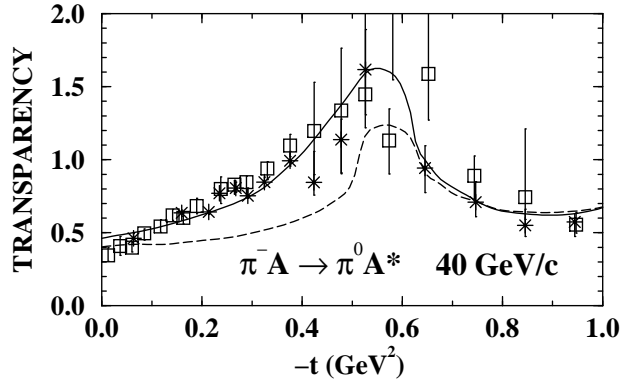

Figure 6: The $t$-dependence of nuclear transparency in the pion charge exchange on nuclei [30]. The solid curve is the prediction of the color dipole model of CT [29], the dashed curve is a prediction of the Glauber model without CT effects.

## 6. CT in quasielastic scattering of electrons $A(e, e'p)$.

### 6.1. The nonobservation of CT effects in the NE18 experiment.

In $A(e, e'p)$ one studies intranuclear attenuation of the struck proton wave produced in $ep$ scattering on a bound proton. The original motivation was that at large $Q^2$ elastic $ep$ scattering is dominated by small size configurations in the proton, hence nuclear attenuation must vanish at large $Q^2$ ([2], it is not yet clear how large $Q^2$ is needed for the short distance dominance [31], though). The by now disproved predictions of the precocious CT are described in [32], the Jülich-Landau group was an exception in predicting a very slow onset of CT [33]. In agreement with [33], no $Q^2$ dependence of nuclear transparency up to $Q^2 \lesssim 7\,\text{GeV}^2$ is seen in the NE18 data [34] shown in Fig. 7.

The path from the above qualitative arguments to the calculations of $T_A$ and to the interpretation of the experimental data is a treacherous one. The ejectile $3q$ state $|E\rangle$ is formed from the proton, when one of the quarks hit by the electron starts moving longitudinally. It is obvious that at short time scales the ejectile has precisely the same transverse size and the same interaction cross section as the free proton [35, 10, 6]. If $l_f \gg R_A$, it is this large size system which undergoes soft intranuclear interactions and is projected out to the observed proton way beyond the nucleus. Weak nuclear effects are possible only at the expense of a conspiracy between form factors of $ep \to ep^*$ electroexcitation which is a hard QCD process, and amplitudes of soft diffractive de-excitation (regeneration of protons) $p^*N \to pN$. The conditions for such a soft-hard conspiracy - the so-called color transparency sum rules [33, 10, 6] - are indeed fulfilled in QCD, but require the very large values of $Q^2$. Refs. [10, 6] explain in detail how the hard QCD mechanism of form factors and CT property of the color dipole cross section lead to a decreasing strength of FSI despite the large initial size of the ejectile state.

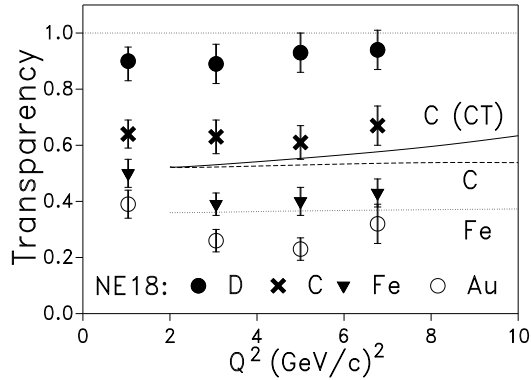

Figure 7: The NE18 data [34] on nuclear transparency for $A(e,e'p)$ as a function of $Q^2$ compared with the Glauber model calculations for the NE18 acceptance [42]. The solid curve shows an estimate of CT effects for $^{12}C(e,e'p)$ in the CT model of the Jülich-Landau group [33, 41, 6].

The point about the range of $Q^2$ in the NE18 experiment is that the kinetic energy of the proton $T_{kin} \approx Q^2/2m_p$ was small, the formation length $l_f$ was short and only in light nuclei and only the low lying excitations $p^*$ could have regenerated the proton wave coherently with the directly electroproduced proton wave. Electroexcitation of the low-lying $\pi N$ continuum which regenerates protons via $(\pi N)N \to pN$ plays an important rôle; alternatively its effect can be described as a depletion of the interaction cross section of the ejectile by $\sim 8$ mb because of stripping off of the pionic field of the proton [36, 10, 6]. (Besides CT, other manifestations of pions in a physical nucleon are the Gottfried sum rule violation and $\bar{u} - \bar{d}$ asymmetry effects in the Drell-Yan production on protons and neutrons [37, 38], the use of these pions as targets for measuring the pion structure function at HERA is discussed in [39]). In certain models (harmonic oscillator interaction between quarks and $\propto r^2$ dipole cross section) the intranuclear evolution of the ejectile state can be described exactly by the path integral technique ([29, 11], for related work see [40]). Experimentally, the $pN \to p^*N$ diffraction excitation amplitudes are small, which suppresses the regenerated proton wave and is a principal reason for the anticipated slow onset of CT [33, 41], in good agreement with the NE18 findings. In Fig. 7 we show an example of the theoretical evaluation of CT effects for the carbon nucleus [41, 6, 42].

The rise of $T_A$ with $Q^2$ signals the onset of CT, but there is a vicious circle in the determination of the absolute value of $T_A$: one is bound [34] to compare the experimentally measured spectral function $S(E_m, \vec{p}_m)$ integrated over the observed range $D$ of the missing energy-momentum $(E_m, \vec{p}_m)$ with the similarly integrated PWIA spectral function: $T_A = \int_D dE_m \, d^3\vec{p} \, S(E_m, \vec{p}_m) / \int_D dE_m \, d^3\vec{p}_m \, S_{PWIA}(E_m, \vec{p}_m)$. The denominator is not measurable, it can only be calculated in theoretical models. The NE18 analysis uses the independent single particle model (ISPM) spectral functions and assumes that $T_A(E_m, \vec{p}_m) = S(E_m, \vec{p}_m)/S_{PWIA}(E_m, \vec{p}_m) = const$. The so deter-mined $T_A$ was further scaled up by 11%, 26% and 32% for the $C, Fe$ and $Au$ targets, respectively, which was meant to correct the ISPM for the correlation effects [34]. One must be careful with such corrections, because the sign of the correction to the PWIA distribution depends on $\vec{p}_m$ [43]. Furthermore, the NE18 analysis neglected distortions of the spectral function for the FSI, which are substantial. The above correlation corrections to $T_A$ are small on the scale of departure of $T_A$ from unity and the NE18 error bars are large, though.

### 6.2. FSI distortions and nuclear transparency

Strong attenuation entails strong distortions which invalidate the PWIA driven assumption $T_A(E_m, \vec{p}_m) = const$. The quantitative theory of distortions is still in its formative stage. It is quite involved and we can not help but cite a bit of the formalism for the simplest case when CT effects are neglected. The best studied quantity is the inclusive missing momentum distribution

$$W(\vec{p}_m) = \frac{1}{(2\pi)^3}\int dE_m S(E_m, \vec{p}_m) = \frac{1}{(2\pi)^3}\int d\vec{r}\,'d\vec{r}\rho(\vec{r},\vec{r}\,')\exp\left[i\vec{p}_m(\vec{r}-\vec{r}\,')\right]. \quad (1)$$

Eq. (1) tells that the $A(e,e'p)$ cross section is a result of quantum interference of amplitudes with different locations of the struck proton. The FSI-distorted one-body density matrix (OBDM) equals $\rho(\vec{r},\vec{r}\,') = \rho_0(\vec{r},\vec{r}\,')\Phi(\vec{r},\vec{r}\,')$, where $\rho_0(\vec{r},\vec{r}\,') = \frac{1}{Z}\sum_n \phi_n(\vec{r})\phi_n^*(\vec{r}\,')$ is the familiar shell model OBDM. If $S(\vec{r},\{\vec{R}\}) = \prod_{j=2}^{A}[1-\theta(Z_j - z)\Gamma(\vec{b}-\vec{B}_j)]$ is the Glauber distortion factor for the fixed configuration of spectators, then $\Phi(\vec{r},\vec{r}\,') = \left\langle S^\dagger(\vec{r}\,',\{\vec{R}\})S(\vec{r},\{\vec{R}\})\right\rangle_{A-1}$, where $\langle...\rangle_{A-1}$ stands for averaging over the spectator configurations $\{\vec{R}\} = \{(\vec{B},Z)\}$. For the simpler case of heavy nuclei the result is [42, 44, 45]

$$\Phi(\vec{r},\vec{r}\,') = \exp\left[-\frac{1}{2}\sigma_{tot}(pN)(1-i\rho)t(\vec{b},z) - \frac{1}{2}\sigma_{tot}(pN)(1+i\rho)t(\vec{b}\,',z')\right.$$
$$\left.+\eta(\vec{b}-\vec{b}\,')\sigma_{el}(pN)t(\frac{1}{2}(\vec{b}+\vec{b}\,'),max(z,z'))\right], \quad (2)$$

where $t(\vec{b},z) = \int_z^\infty dz_1 n_A(\vec{b},z_1)$ is a partial optical thickness, $\Gamma(\vec{b})$ is the profile function of $pN$ scattering, $\rho$ is the $Re/Im$ ratio for the $pN$ scattering amplitude and

$$\eta(\vec{b}) = \frac{\int d^2\vec{\Delta}\Gamma^*(\vec{b}-\vec{\Delta})\Gamma(\vec{\Delta})}{\int d^2\vec{\Delta}|\Gamma(\vec{\Delta})|^2} = \frac{1}{\pi\sigma_{el}(pN)}\int d^2\vec{q}\frac{d\sigma_{el}(pN)}{dq^2}\exp(i\vec{q}\vec{b}) = \exp\left[-\frac{\vec{b}^2}{4b_0^2}\right]. \quad (3)$$

The DWIA factorization approximation $\Phi(\vec{r},\vec{r}\,') = \left\langle S^\dagger(\vec{r}\,',\{\vec{R}\})\right\rangle_{A-1}\left\langle S(\vec{r},\{\vec{R}\})\right\rangle_{A-1}$ reproduces the first two terms in the exponent of (2) but misses the crucial term $\propto \eta(\vec{b}-\vec{b}\,')$, the rôle of which was elucidated in recent works [46, 41, 42, 47, 43]. In the cluster expansion of $S^\dagger S$, it derives from the terms $\propto \Gamma^\dagger\Gamma$ which describe the quantal interference effect when both trajectories which enter the calculation of the FSI-modified OBDM interact with the same spectator, see Fig. 8. In view of (3), the latter can be identified with incoherent elastic rescatterings of the struck proton on the spectator nucleons.

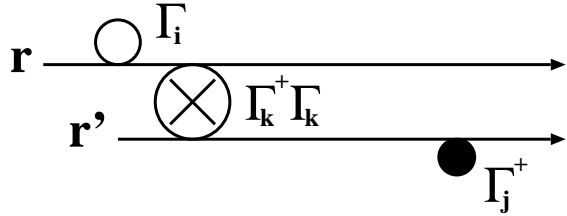

Figure 8: Open, closed and crossed circles denote the $\Gamma, \Gamma^\dagger$ and $\propto \Gamma^\dagger \Gamma$ interactions in the calculation of the FSI-distorted one body density matrix for $A(e, e'p)$.

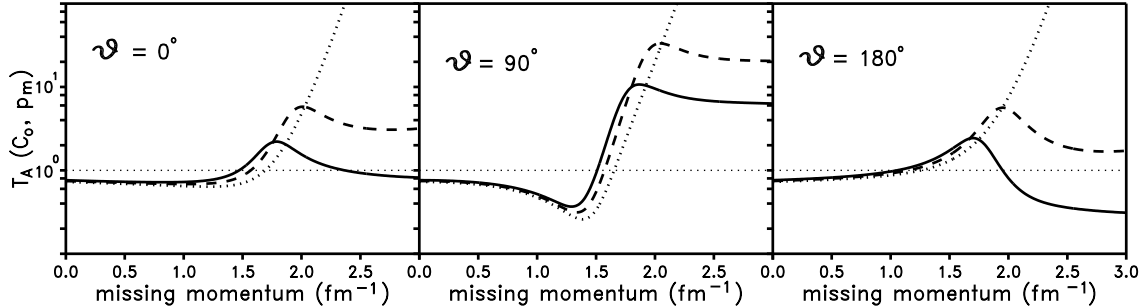

Figure 9: Nuclear transparency $T_A(C_0, \vec{p}_m)$ for $^4He(e, e'p)$ for the hard core ($C_0 = 1$, the solid curve) ), soft core ($C_0 = 0.5$, the dashed curve) and vanishing ($C_0 = 0$, the dotted curve) correlation [43] for the parallel, transverse and antiparallel kinematics.

Apart from breaking the DWIA factorization, the crucial property of $\eta(\vec{b} - \vec{b}')$ is a rapid dependence on $\vec{b} - \vec{b}'$ on the short scale $b_0$, which is close to the radius $r_c$ of short range correlations. In transverse kinematics, this leads to strong distortions of $W(\vec{p}_m)$ which at large $\vec{p}_m$ completely take over the usually discussed short range correlation effects [47, 43] (Fig. 9). The relevant large parameter is $R_A/r_c \gg 1$. Fig. 9 clearly shows a weak sensitivity of nuclear transparency $T_A(p_m)$ to short range correlations up to $p_m \lesssim 1.7$-$1.8$ fm$^{-1}$ and even beyond, barring the case of $C_0 = 0$ which is not realistic at large $p_m$. Notice that $T_A(p_m) \gg 1$ at large $p_m$ in transverse kinematics. The approximation [44, 45] of putting $\eta(0)$ in the exponent of the $\Phi(\vec{r}, \vec{r}')$ misses these strong distortions. For heavy nuclei, one can derive the multiple elastic rescattering expansion for the $p_{m,z}$-integrated $p_{m\perp}$ distribution [42], a simplified form of this expansion for large-$p_m$ is given in [46, 41].

In parallel kinematics, the principal distortions are the strong forward-backward (F-B) asymmetry (see Fig. 9) caused by the effective shift of the missing momentum by $k_1 = \frac{1}{2}\sigma_{tot}(pN)\alpha_{pN}\langle n_A \rangle$ [41, 42] and FSI-induced large $p_m$ tails of $W(\vec{p}_m)$. The latter derives [47, 42] from a nonanalytic $z - z'$ dependence of the term $\propto \eta(\vec{b} - \vec{b}')$ in the exponent of (2) because of the nonanalytic function $max(z, z')$, which is of a purely quantum origin and tells that the $\propto \Gamma^\dagger \Gamma$ interaction is possible only on the longitudinally overlapping parts of the two trajectories in the calculation of the FSI-modified OBDM. This effect is missed in the conventional DWIA. Fig. 9 clearly shows that even in light nuclei the distortions are strong and evaluations of $T_A$ under the assumption of $T_A(E_m, \vec{p}_m) = const$ can lead astray.

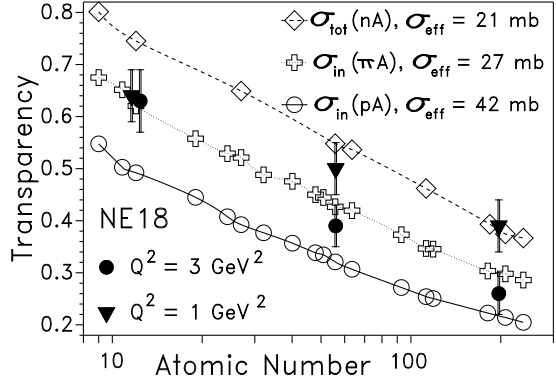

Figure 10: Target mass dependence of nuclear transparency as measured in $nA$ [50], $\pi A$ and $pA$ [49] and NE18 $A(e, e'p)$ scattering experiments.

6.3. How strong is FSI observed in $A(e, e'p)$ scattering?

Nuclear transparency for the fully $(E_m, \vec{p}_m)$-integrated cross section equals [33]

$$T_A = \int dz d^2\vec{b}\, \frac{n_A(\vec{b}, z)}{A} \exp\left[-\sigma_{in}(pN) t(\vec{b}, z)\right] = \frac{\int d^2\vec{b} \left\{1 - \exp\left[-\sigma_{eff} t(\vec{b}, \infty)\right]\right\}}{A \sigma_{in}(pN)}. \quad (4)$$

The emergence of $\sigma_{eff} = \sigma_{in}(pN)$ is very natural: elastic rescatterings only deflect, do not absorb, the struck proton. (One must be careful with such probabilistic arguments, though [42]). The result (4) is unique for the lack of sensitivity to the form of $S_{PWIA}(E_m, \vec{p}_m)$. The numerator in (4) is precisely the $pA$ absorption cross section. For $T_A(\vec{p}_m \sim 0)$ one finds to a crude approximation the same formula (4) but with $\sigma_{eff} = \sigma_{tot}(pN)$. If the experimental $p_{m\perp}$ integration range is sufficiently broad, then one must find $\sigma_{eff}$ which interpolates between $\sigma_{tot}(pN)$ and $\sigma_{in}(pN)$ [46, 41, 42]; the NE18 analysis cites an estimate $\sigma_{eff} \approx 0.68 \sigma_{tot}(pN)$ [34]. As a matter of fact, $T_A$ has been measured in the hadron-nucleus interactions for a very broad range of $\sigma_{eff}$: at $\sigma_{eff} = \sigma_{tot}(hN), \sigma_{in}(hN), \frac{1}{2}\sigma_{tot}(hN)$ in the measurements of $\sigma_{in}(hA), \sigma_{abs}(hA), \frac{1}{2}\sigma_{tot}(hA)$, respectively [48], and in Fig.10 we show $T_A$ from these nuclear cross section data ($pA, \pi A$ [49], $nA$ [50]). We conclude that the observed strength of FSI is consistent with that expected for the free-nucleon cross section: $\sigma_{tot}(pN) \approx 33\,\mathrm{mb}$, $\sigma_{in}(pN) \approx 20\,\mathrm{mb}$ at $T_{kin} = 0.53\,\mathrm{GeV}$ ($Q^2 = 1\,\mathrm{GeV}^2$) and $\sigma_{tot}(pN) \approx 45\,\mathrm{mb}$, $\sigma_{in}(pN) \approx 25\,\mathrm{mb}$ at $T_{kin} = 1.6\,\mathrm{GeV}$ ($Q^2 = 3\,\mathrm{GeV}^2$). Notice the anticorrelation of the energy dependence of the observed $T_A$ in Fig.7 with the energy dependence of $\sigma_{tot}(pN), \sigma_{in}(pN)$. One must bear in mind that $T_{kin} \sim 0.5\,\mathrm{GeV}$ is still too low for the Glauber calculations to be highly accurate, there are also uncertainties with the above cited correlation corrections to the NE18 values of $T_A$.

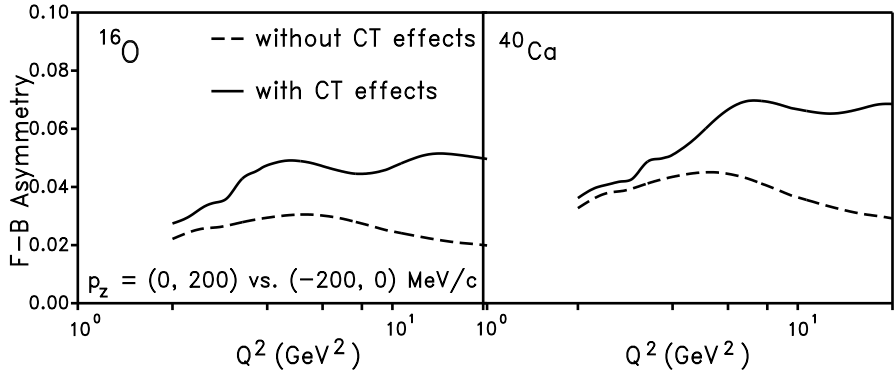

Figure 11: The predicted [54] F-B asymmetry $A_{FB} = (N_+ - N_-)/(N_+ + N_-)$, where $N_\pm$ are numbers of events in the longitudinal missing momenta intervals $(0, 200)$MeV/c and $(-200, 0)$MeV/c, respectively. The solid curve includes CT effects, the dashed curve is the Glauber theory result without CT effects.

6.4. *CT and forward-backward asymmetry.*

By the kinematics of $ep \to eX$, the elastic scattering $X = p$ and electroexcitation $X = p^*$ proceed on target protons which have the longitudinal Fermi momenta differing by $\Delta k_z = m_p(m_X^2 - m_p^2)/Q^2$. Consequently, zooming at proper $p_{m,z}$, one can either enhance or suppress the outgoing proton wave regenerated from the electroproduced $p^*$ states, which leads to the F-B asymmetry $A_{FB}$ of $W(\vec{p}_m)$ [51, 52, 53]. However, a strong background to this signal of CT comes (Fig. 11, see also Fig. 9) from the FSI-induced asymmetry due to the real part of the $pN$ scattering amplitude [41, 6, 42]. Hopefully, one can disentangle the CT effect using its $Q^2$ dependence.

6.5. *Is the normal polarization a CT observable?*

The normal polarization $P_n$ of the struck proton in $A(e, e'p)$ only comes from FSI. Will then the $P_n$ vanish in the CT regime? The difficult experiments on the $Q^2$ dependence of $P_n$ are planned at CEBAF. We notice that the spin-orbit interaction is reggeon-exchange dominated. Ref. [29] predicted a very strong CT-induced enhancement of the normal polarization of the recoil nucleons in the CEX of pions (Section 5). In $A(e, e'p)$ too, CT effects can enhance the reggeon exchange amplitudes at finite transverse missing momentum, and the normal polarization of struck protons will be enhanced even when CT suppresses the dominant FSI for the pomeron exchange central interaction. This difference between the reggeon and pomeron exchange was not considered in [55]. The same holds of the other spin observables, the status of $P_n$ and of other spin correlations as a CT observable is questionable, more theoretical work is needed.

## 6.6. CT and electroproduction of pions $ep \to e'\pi p$

This is a unique reaction for measuring the pion electromagnetic form factor at large spacelike $Q^2$. The normalization of the cross section is subject to the pion-nucleon FSI, which goes away at large $Q^2$ for the CT effects [56]. In terms of the Fock state expansion of the dressed proton, the underlying reaction mechanism is electroexcitation of the pion-nucleon Fock state of the dressed proton, this $\pi N$ state acts as a mini-nucleus and for the small size of the $\pi N$ state one expects the precocious CT.

## 7. Quasielastic scattering of protons $A(p, 2p)$.

$A(p, 2p)$ scattering on bound nucleons at $\approx 90^0$ in the $pp$ c.m.s was studied at BNL AGS [57]. The beam momenta $\vec{p}_1$ were 6, 10 and 12 GeV, the momentum $\vec{p}_3$ of one ejected proton and the emission angle of the second ejectile of momentum $\vec{p}_4$ were measured, the Fermi momentum $\vec{p}$ of the target proton ($\vec{p} = -\vec{p}_m$) was kinematically reconstructed. The values of longitudinal momentum $p_z$ up to $|p_z| = 0.3$ GeV/c were used to stretch the range of the effective c.m.s. energy squared $s_{eff} \approx s(1 - \frac{p_z}{m_p})$. The data were analyzed assuming sort of a PWIA factorization $d^2\sigma_A/dtdp_z = ZT_A W_z(p_z)d\sigma_N(s_{eff}, t)/dt$. The unknown $p_z$ distribution $W_z(p_z)$ was assumed to be identical to the distribution $W_y(p_y)$ of the transverse out-of plane momentum $p_y$, measured in the same experiment (hereafter $W_{z,y}(p)$ are normalized to unity). This assumption can not be viable: what we learnt from $A(e, e'p)$ scattering (for instance, see Fig. 9) is that for strong distortions $W_z(p) \neq W_y(p)$ [47, 42, 43]. The FSI driven difference between the $W_z(p)$ and $W_y(p)$ can only be enhanced by the presence of three strongly interacting protons in $A(p, 2p)$ and, moreover, the $A(p, 2p)$ and $A(e, e'p)$ must yield very different distributions $W(\vec{p})$. Furthermore, the very recent accurate quantum-mechanical theory of $A(p, 2p)$ reveals [58] a nontrivial and numerically important interference between the initial state interactions (ISI) of the projectile proton and FSI of ejected protons, which generalize the interaction $\propto \Gamma^\dagger\Gamma$ in $A(e, e'p)$.

In $A(p, 2p)$ one must evaluate the FSI-modified OBDM with the distortion factor

$$\Phi(\vec{r}, \vec{r}\,') = \left\langle S_1^\dagger(\vec{r}\,', \{\vec{R}\})S_3^\dagger(\vec{r}\,', \{\vec{R}\})S_4^\dagger(\vec{r}\,', \{\vec{R}\})S_4(\vec{r}, \{\vec{R}\})S_3(\vec{r}, \{\vec{R}\})S_1(\vec{r}, \{\vec{R}\}) \right\rangle_{A-1}. \tag{5}$$

Here $S_1$ describes the ISI of the projectile and $S_{3,4}$ describe the FSI of ejectiles. Apart from the interactions $\Gamma_i, \Gamma_i^\dagger$ (the open and full circles in Fig.12a) and $\propto \Gamma^\dagger\Gamma$ (the crossed circles in Fig.12b) familiar from $A(e, e'p)$, the cluster expansion of (5) gives rise to the host of unusual interactions shown partly in Fig. 12b: $\propto \Gamma_3\Gamma_4 + h.c.$ (the encircled full triangles) describe simultaneous FSI of both ejectiles with the same spectator, $\propto \Gamma_1^\dagger\Gamma_{3,4} + h.c.$ (the encircled full stars) describe the still more unusual quantal interference between ISI of the projectile and FSI of ejectiles on the longitudinally overlapping parts of trajectories as shown in Fig. 12b, there are still higher order ternary and quartic terms in (5). These unusual interactions are short ranged in $\vec{b} - \vec{b}\,'$ and nonanalytic in $z - z'$; they are not weak and produce strong distor-tions at large $\vec{p}$; their effects are completely missed in the conventional DWIA, which amounts to summing up only the diagrams of Fig. 12a. Even in the fully $\vec{p}$ integrated cross section, the effect of these unusual interactions does not vanish  The distortion effects for $O^{16}(p,2p)$ are shown in Fig. 13 in the form of $R(\vec{p}) = W(\vec{p})/N(\vec{p})$. The very strong enhancement at large $p_y$ was indeed observed in the BNL experiment [57]. When these distortions are taken into account, the theory yields $T_A$ which has a very steep $p_z$ dependence [58], which is shown in Fig. 14 and is similar to the BNL AGS observations. and there is no room for the semiclassical attenuation of the projectile and ejectiles with $\sigma_{in}(pN)$ [44, 59] like the one one recovered in the simpler $A(e,e'p)$ case, see Eq. (4). The $p_z$-dependence of $T_A$ also comes from the forward-backward asymmetric CT effect discussed in Section 6.5; the effect was evaluated in [60] and is weaker than the ISI/FSI distortion effect. The first quantum-mechanical evaluation of the energy dependence of CT effects was reported in [29].

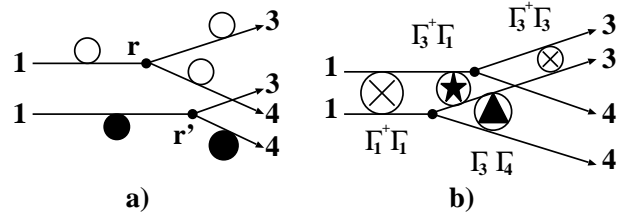

Figure 12: Interactions in the calculation of the ISI-FSI distorted OBDM for $(p,2p)$.

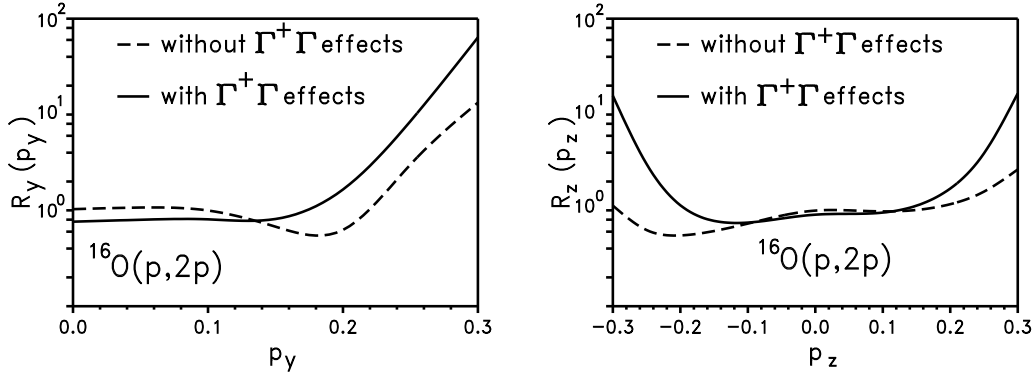

Figure 13: Distortions of the transverse $(p_y)$ and longitudinal $(p_z)$ missing momentum distributions in $^{16}O(p,2p)$ scattering [58].

Regarding the model dependence of $T_A$, one must be aware that the energy dependence of the $90^\circ$ $pp$ cross section is very steep, $\sim 1/s^{10}$. Consequently, the usually discussed binding corrections $\Delta m \sim$-0.03-0.05 GeV can significantly (by the factor $\sim$1.3-1.7) enhance the nuclear cross section. At $p_y \sim 0.3$GeV the constructive [47] interference of the ISI and FSI with short range correlation effects can be substantial. The $\propto \Gamma_3 \Gamma_4$ and $\propto \Gamma_1^\dagger \Gamma_{3,4}$ interference effects are sensitive to the radius of the $90^\circ$ $pp$ scattering, which can be large if the Landshoff mechanism contributes [61]. Finally, the observed energy dependence of $T_A$ anticorrelates with the oscillatory behavior of $s^{10}d\sigma_{pp}/dt$ [62]. We are lead to the conclusions that, i) PWIA driven analysis of the $A(p,2p)$ scattering is not viable and must be discarded, ii) ISI-FSI distortions [58] is the first mechanism which gave rise to the irregular $p_z$ dependence of the form observed experimentally, iii)$(p,2p)$ scattering is a reaction the interpretation of which is too involved for unambiguous quantitative conclusions on CT effects.

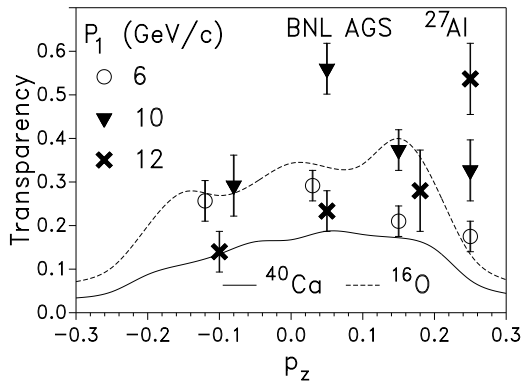

Figure 14: Glauber model predictions [58] for nuclear transparency in $A(p,2p)$ scattering vs. the longitudinal Fermi momentum $p_z$ of the target proton. The data points are for $^{27}Al$ target [57], the dashed and solid theoretical curves are for closed shell nuclei $^{16}O$ and $^{40}Ca$ and for the beam momentum $p_1 = 10$ GeV.

## 8. Conclusions.

CT ideas are on the solid QCD footing and we have a good understanding of the onset of CT effects in different exclusive reactions. Electroproduction of vector mesons on nucleons and nuclei is the best understood reaction and the experiments provide a solid evidence for CT. Here the important new result is a direct experimental evidence for the fundamental ingredient of the theory of CT, namely the CT property of the color dipole cross section. There are many interesting new predictions for production of the radially excited $2S$ mesons which can be tested at CEBAF, ELFE, HERA. The theoretical interpretation of quasielastic $A(e, e'p)$, $A(p, 2p)$ scattering, as of any other reaction which is sensitive to the momentum of the Fermi motion of the bound target nucleon, is less straightforward and is subject to the quantitative understanding of the ISI-FSI distortions. In this field too we witness a rapid development. It is fair to say that the subject of CT is entering its maturity.

**Acknowledgments:** Thanks are due to S.Jeschonnek and H.Holtmann for their assistance in preparation of this talk. This work was partly supported by the INTAS grant No. 93-239.